\documentclass[aps,prc,twocolumn,groupedaddress,showpacs]{revtex4-1}
\usepackage{graphicx}
\usepackage{dcolumn}
\usepackage{bm}
\usepackage{amssymb}
\usepackage{longtable}

\begin{document}

\title{Reduced $E$1 $S$-factor of $^{12}$C($\alpha$,$\gamma_0$)$^{16}$O}
\author{M. Katsuma}
\email[]{mkatsuma@gmail.com}
\affiliation{Advanced Mathematical Institute, Osaka City University, Osaka 558-8585, Japan}

\date{\today}

\begin{abstract}
  The astrophysical $S$-factor of $E$1 transition for $^{12}$C($\alpha$,$\gamma_0$)$^{16}$O is discussed in the $R$-matrix theory.
  The reduced $\alpha$-particle widths of the 1$^-_1$ ($E_x= 7.12$ MeV) and 1$^-_2$ ($E_x= 9.59$ MeV) states are extracted from the result of the potential model.
  The formal parameters are obtained without the linear approximation to the shift function.
  The resultant $E$1 $S$-factor is not strongly enhanced by the subthreshold 1$^-_1$ state if the channel radius is 4.75 fm.
  The calculated $\beta$-delayed $\alpha$-particle spectrum of $^{16}$N and the $p$-wave phase shift of $\alpha$+$^{12}$C elastic scattering are also found to be consistent with the previous studies.
  The small channel radius leads to the low penetrability to the Coulomb barrier, and it makes the reduced $E$1 $S$-factor below the barrier.
  Owing to the large reduced width from the molecular structure, the $R$-matrix pole of the 1$^-_2$ state is shifted in the vicinity of 1$^-_1$.
  The proximity of the two poles suppresses the interference between the states.
  The transparency of the $\alpha$+$^{12}$C system appears to be expressed as the shrinking strong interaction region.
\end{abstract}

  
\maketitle

\section{Introduction}

  The C/O ratio at the end of the helium burning phase determines the fate of stars, and it affects the various type of the nucleosynthesis after the helium burning phase.
  The C/O ratio is controlled by the $^{12}$C($\alpha$,$\gamma$)$^{16}$O reaction.
  So, the $^{12}$C($\alpha$,$\gamma$)$^{16}$O reaction is thought to be a key reaction of the nucleosynthesis of heavy elements.
  However, the determination of the reaction rates for this reaction has the experimental difficulties.
  The most important energy corresponding to the helium burning temperature is $E_{c.m.}= 300$ keV \cite{Rol88}.
  ($E_{c.m.}$ is the center-of-mass energy of the $\alpha$+$^{12}$C system.)
  This reaction energy is too low to reach by the present laboratory technology.
  When the reaction rates are estimated, the low-energy cross section is extrapolated from the available experimental data by the theoretical model to cope with the unknown tiny cross section due to the Coulomb barrier.

  In the analyses, the $\alpha$-particle width of the subthreshold 1$^-_1$ state at $E_x= 7.117$ MeV in $^{16}$O has been believed to be essential to determine the $E$1 cross section at $E_{c.m.}= 300$ keV.
  ($E_x$ denotes the excitation energy.)
  The 1$^-_1$ state is described by the particle-hole excitation in the shell model (e.g. \cite{Zuk68}), and it is located at the excitation energy just below the $\alpha$-particle threshold.
  Because it is a bound state, the 1$^-_1$ state does not have a decay width for $\alpha$-particle emission, but a reduced width describing the probability of the $\alpha$-particle at the nuclear surface.
  The reduced width is obtained from the $\alpha$-particle spectroscopic factor or the asymptotic normalization constant (ANC) \cite{Muk95,Muk99}.
  To estimate it experimentally, the indirect measurements (e.g. \cite{Oul12,Bel07,Bru99}), including the $\beta$-delayed $\alpha$-particle spectrum of $^{16}$N (e.g. \cite{Azu94,Azu97,Buc96a,Buc06,Zha93a,Tan07,Tan10}), have been performed recently.
  The direct measurements of $\gamma$-ray angular distribution have also been performed in \cite{Kun01,Ass06,Ham05a,Ham05b,Pla05,Pla12,Oue96,Mak09}.
  In addition, the cascade transition through the 1$^-_1$ state \cite{Red87,Sch12,Sch11} has been measured, and the $\alpha$+$^{12}$C elastic scattering has been investigated in \cite{Pla87,Tis02,Tis09}.
  In spite of all these experimental efforts, the reduced width and the $E$1 cross section have not been understood satisfactorily yet.

  The surface probability of $\alpha$-particle originates from a component of the $\alpha$+$^{12}$C configuration in the 1$^-_1$ state.
  Especially, the 1$^-_2$ state ($E_x\approx 9.585$ MeV) is described by the $\alpha$+$^{12}$C cluster structure \cite{Kat13,Kat14a,Kat10b,Fuj80,Suz76a,Suz76b}, so that the coupling between 1$^-_1$ and 1$^-_2$ is thought to play an important role in the low-energy extrapolation of the $E$1 cross section.
  If the strong interference between two states happens, the $E$1 cross section will be consequently enhanced by the 1$^-_1$ state at low energies.
  The $R$-matrix theory (e.g.~\cite{Des,Ang00,Des04,Tho09,Hum91,Lan58,Tho51}) is used as a popular method to describe the state coupling in the $^{12}$C($\alpha$,$\gamma$)$^{16}$O reaction.

  It is, however, pointed out that the $\alpha$+$^{12}$C system has an inherent problem on the definition of the channel region in the $R$-matrix method \cite{Kat08,Kat10a}.
  The excitation function of $\alpha$+$^{12}$C elastic scattering below $E_{c.m.}=5$ MeV is expressed as potential scattering without absorption \cite{Kat10b}.
  This means that the $\alpha$+$^{12}$C system is almost completely transparent in the entire radial region.
  On the other hand, the $R$-matrix theory assumes the spherical strong interacting region with the sharp-cut edge.
  The compound nucleus is formed inside the sphere.
  The $\alpha$ and $^{12}$C nuclei are well-identified outside the sphere.
  In general, the channel radius $a_c$ corresponds to the radius of the strong absorptive region or strongly interacting region.
  Therefore, the value of $a_c$ is not defined on firm ground for the $\alpha$+$^{12}$C system.
  None the less, one may think that the $R$-matrix method works effectively if the boundary condition is adjusted so that the $\alpha$+$^{12}$C configuration becomes dominant in the $^{16}$O nucleus.
  The large dimensionless width to the Wigner limit is expected from the dominant $\alpha$+$^{12}$C configuration.
  This gives the large reduced width at a short $a_c$.
  Meanwhile, the large reduced width for 1$^-_2$ has been reported to lead a defect unexpectedly in the linear approximation of the resonance parameters \cite{Des}.
  From the imperfection in the available range of $a_c$, one may surmise that the popular value of $a_c$ is not better in the optimization.
  In the calculable $R$-matrix method (e.g.~\cite{Des10b,Duf08}), the internal wavefunction is generated by the variational method in order to reveal couplings with other degrees of freedom.

  From the transparency of the system, the weak coupling between $\alpha$+$^{12}$C and other configurations can be expected in $^{16}$O.
  In my previous articles \cite{Kat08,Kat10a}, the reduced $E$1 $S$-factor has been predicted with the potential model.
  The $S$-factor is used conventionally, instead of the low-energy cross section, to compensate for the rapid drop below the Coulomb barrier.
  
  In the present article, I illustrate the reduced $E$1 $S$-factor of $^{12}$C($\alpha$,$\gamma_0$)$^{16}$O with the $R$-matrix theory.
  The $p$-wave phase shift of $\alpha$+$^{12}$C elastic scattering and the $\beta$-delayed $\alpha$-particle spectrum of $^{16}$N are also calculated.
  The input reduced $\alpha$-particle widths for 1$^-_1$ and 1$^-_2$ are extracted from the wavefunction in the potential model \cite{Kat13,Kat14a,Kat10b,Kat08,Kat10a,Kat12,Kat15}.
  In addition, the higher-order correction to the linear approximation of the resonance parameters is examined because the large reduced width is adopted \cite{Des,Tho09}.
  The purpose of the present article is to exemplify the reduced $E$1 $S$-factor at $E_{c.m.}=300$ keV by the $R$-matrix method and to assess the sensitivity to the channel radius.

  In the following section, I describe the difference between the present model and the widely used $R$-matrix method.
  In Sec.~III, I illustrate an example of the reduced $E$1 $S$-factor.
  I also show the corresponding results of the $\beta$-delayed $\alpha$-particle spectrum of $^{16}$N and the $p$-wave phase shift for $\alpha$+$^{12}$C elastic scattering.
  After discussing the sensitivity to the channel radius, I summarize the present article in Sec.~IV.

\section{Resonance parameters in $R$-matrix}

  I use the conventional $R$-matrix method in the present article.
  In this section, let me describe two differences from the previous $R$-matrix method of the $^{12}$C($\alpha$,$\gamma_0$)$^{16}$O reaction.
  One is the estimation of the reduced $\alpha$-particle width, and the other is the correction for the linear approximation of the resonance parameters.
  The $R$-matrix theory used in the present article is described in Appendix, and the detail can be found in \cite{Des,Ang00,Des04,Tho09,Hum91,Lan58,Tho51}.

\begin{figure}[t]
  \centerline{
    \begin{tabular}{c}
      \includegraphics[width=0.65\linewidth]{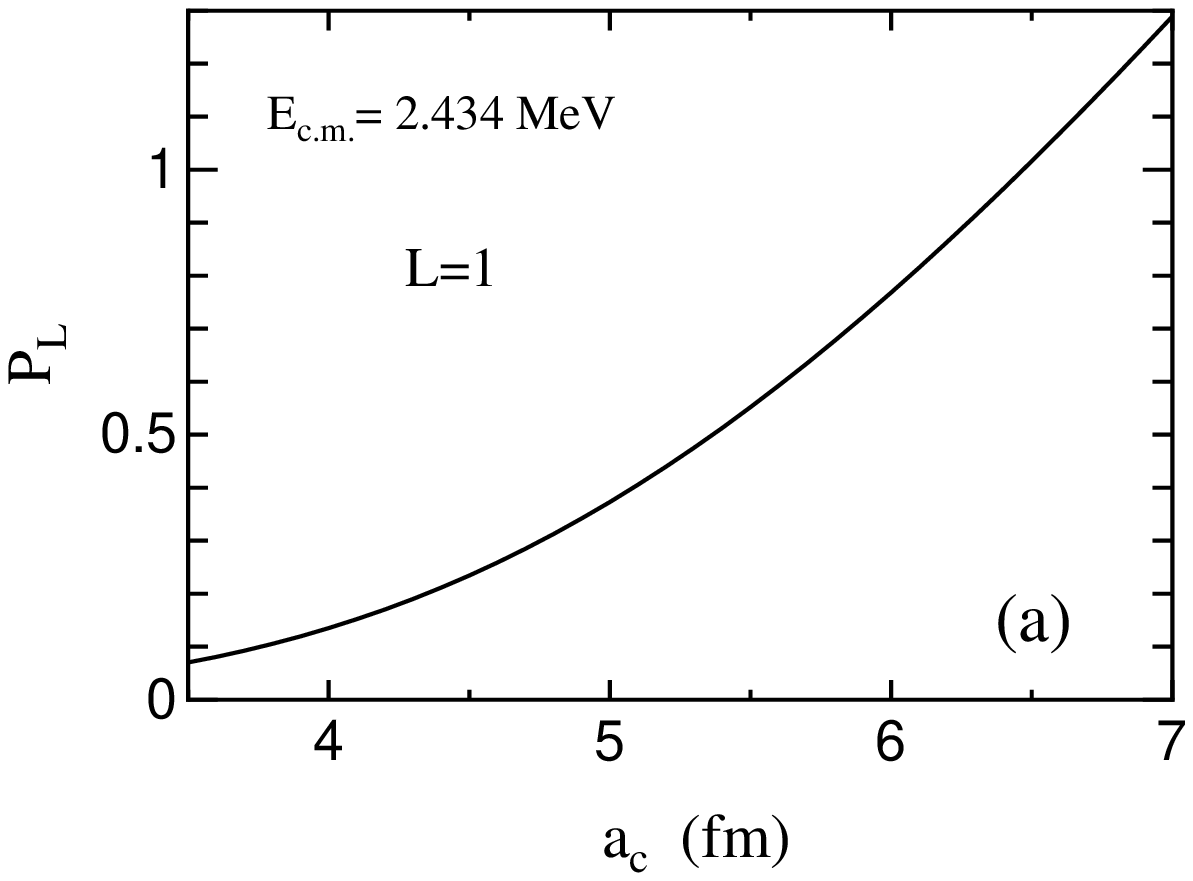} \\
      \includegraphics[width=0.65\linewidth]{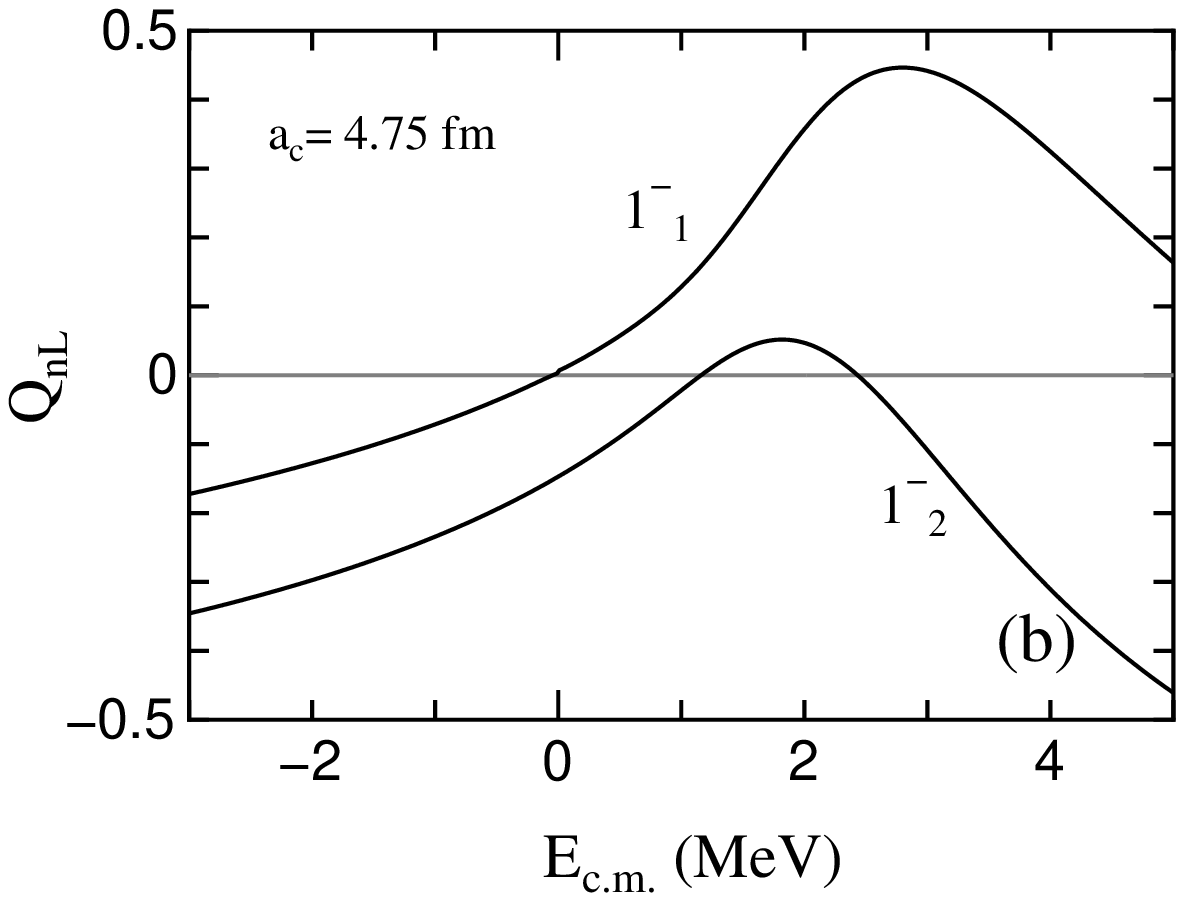} \\
    \end{tabular}
  }
  \caption{\label{fig:coulf}
    Utilized Coulomb function.
    (a) Penetration factor $P_L$ defined in Eq.~(\ref{eq:pene}) as a function of the channel radius $a_c$.
    The solid curve is obtained with $E_{c.m.}=2.434$ MeV and $L=1$.
    The large $a_c$ makes the high penetrability of the barrier, i.e. it corresponds to the reduction of the Coulomb barrier.
    (b) Higher-order correction term $Q_{nL}(E_{c.m.},a_c)$ in Eq.~(\ref{eq:q-ecm}) with $a_c=4.75$ fm.
    The solid curves are calculated for the 1$^-_1$ and 1$^-_2$ states.
    $Q_{n1}=0$ is used in the linear approximation to the shift function.
  }
\end{figure}

The four 1$^-$ states, 1$^-_1$, 1$^-_2$, 1$^-_3$ ($E_x= 12.44$ MeV), and 1$^-_4$ ($E_x= 13.09$ MeV) \cite{Til93}, are included in the $R$-matrix calculation.
The reduced $\alpha$-particle width $\gamma_{n L}$ is labeled with $n$ and $L$.
$L$ is the angular momentum of the relative motion between $\alpha$+$^{12}$C.
$n$ is the ordinal number of the state with $L$ in order of the excitation energy.
The reduced $\alpha$-particle width for the subthreshold 1$^-_1$ state is obtained from
  \begin{eqnarray}
    \gamma^2_{1 1} &\approx& \frac{\hbar^2}{2\mu a_c}
    C^2 \left|\,W_{-\eta,3/2}(2k_b a_c)\,\right|^2,
    \label{eq:sub-g}
  \end{eqnarray}
where $C$ denotes ANC, $C^2 = 5.0\times10^{28}$ fm$^{-1}$ \cite{Kat08,Oul12,Bel07,Bru99}.
$W$ is the Whittaker function.
$\eta$ is the Sommerfeld parameter.
$k_b$ is the wave number of the bound state, $k_b= \sqrt{2|E_b|\mu/\hbar^2}$; $\mu$ is the reduced mass; $E_b$ is the binding energy.
In the conventional $R$-matrix method, the internal wavefunction is not calculated by solving the Schr\"odinger equation numerically.
If $a_c$ is small, the $\gamma_{1 1}$ is more appropriately given as
  \begin{eqnarray}
    \gamma^2_{1 1} &=& \frac{\hbar^2}{2\mu a_c}
    S_\alpha \left|\,\varphi_B(k_b a_c)\,\right|^2,
    \label{eq:sub-g-sp}
  \end{eqnarray}
where $S_\alpha$ is the spectroscopic factor.
$\varphi_B$ is the bound state wavefunction generated from the potential \cite{Kat08} reproducing the $\alpha$-particle separation energy.
It is noted that the reduced width is estimated only from the wavefunction at $r=a_c$.
$\gamma^2_{11}$ is independent of the radial node of the wavefunction in the internal region.
I use this value as a guide of $\gamma^2_{11}$, instead of Eq.~(\ref{eq:sub-g}).
The same discussion can be made with Eq.~(\ref{eq:sub-g}).
The 1$^-_2$ resonant state is a member of the $\alpha$+$^{12}$C rotational bands.
The reduced $\alpha$-particle width $\gamma_{2 1}^2$ is obtained from the wavefunction of potential scattering at $r=a_c$ \cite{Kat08,Kat10b}, and it is given in the similar expression to Eq.~(\ref{eq:sub-g-sp}) with $S_\alpha=1$.
In the calculation, the asymptotic form of the scattering wavefunction is defined as
  \begin{eqnarray}
    \chi_L^{\rm PM}(k_i,r)
    &\rightarrow&\!
    e^{i\delta^{\rm N}_L}\!\!\left[ F_L(k_i r) \cos\delta^{\rm N}_L +G_L(k_i r) \sin\delta^{\rm N}_L \right],
    \label{eq:chi_pm}
  \end{eqnarray}
where $F_L(k_i r)$ and $G_L(k_i r)$ are the regular and irregular Coulomb wave functions, respectively.
$\delta^{\rm N}_L$ is the nuclear phase shift.
$k_i$ is the wave number, $k_i = \sqrt{2\mu E_{21}/\hbar^2}$.
$E_{nL}$ is the observed resonance energy.
$E_{21}$ is adjusted within $E_{21}= (2.423 \pm 0.011)$ MeV.
The normalized scattering wave is give by $\varphi_S(k_i r)=\sqrt{2/(\pi \hbar v)}\,\chi_L^{\rm PM}(k_i,r)$.
$v$ is the velocity of the relative motion between $\alpha$ and $^{12}$C nuclei.
The observed $\alpha$-particle width $\Gamma_{n L}$ is defined in
  \begin{eqnarray}
    \Gamma_{n L} &=& 2 P_L(E_{nL},a_c) \,\gamma_{n L}^2 ,
    \label{eq:ga}
  \end{eqnarray}
where $P_L$ is the penetration factor defined in Eq.~(\ref{eq:pene}).
The example of the penetration factor is shown in Fig.~\ref{fig:coulf}(a) as a function of $a_c$.
The $a_c$ modulates the barrier penetrability, that becomes low when $a_c$ is small as if the Coulomb barrier is high.
In order to take account of the contribution from 1$^-_3$ and 1$^-_4$ at the high excitation energies, $\gamma_{31}$ and $\gamma_{41}$ are included within $\Gamma_{31}= (92\pm8)$ keV and $\Gamma_{41}=(45\pm18)$ keV \cite{Til93}.
To examine the state, the dimensionless width $\theta_{n L}$ is defined as
  \begin{eqnarray}
    \theta_{n L}^2 &=& \frac{\gamma_{n L}^2}{\gamma_W^2},
    \label{eq:towig}
  \end{eqnarray}
where $\gamma_W$ denotes the Wigner limit, $\gamma_W^2=3\hbar^2/(2\mu a_c^2)$.

The resonance parameters in the $R$-matrix theory are different from those in the Breit-Wigner type of the experimental resonance.
The {\it formal} resonance energy $\tilde{E}_{n L}$ and {\it formal} reduced width $\tilde{\gamma}_{n L}$ of the $n$th pole in the $R$-matrix theory are defined in the linear approximation for the single pole as
  \begin{eqnarray}
    \tilde{E}_{n L} &=& E_{n L} + \tilde{\gamma}_{n L}^2 \Delta_L(E_{n L}, a_c),
    \label{eq:formal-e}\\
    \tilde{\gamma}_{n L}^2 &=& \frac{\gamma_{n L}^2}{1-\gamma_{n L}^2 \Delta_L^\prime(E_{n L}, a_c)},
    \label{eq:formal-g}
  \end{eqnarray}
where $\Delta_L$ is the shift function in Eq.~(\ref{eq:shift}), and $\Delta_L^\prime(E_{c.m.}, a_c)=d\Delta_L/dE_{c.m.}$.
To obtain Eqs.~(\ref{eq:formal-e}) and (\ref{eq:formal-g}), $\Delta_L(E_{c.m.}, a_c)$ is expressed linearly at $E_{nL}$.
By this approximation, both formal and observed parameters are independent of $E_{c.m.}$ in \cite{Lan58}.
Although it may be widely used in the $R$-matrix code, the linear approximation is valid only when the reduced width is narrow.
In the present article, the 1$^-_2$ state is expected to have the large reduced width due to the $\alpha$+$^{12}$C molecular state.
So, this state cannot be treated in the conventional procedure \cite{Des}.
To treat the 1$^-_2$ state accurately, I introduce the higher-order correction to Eq.~(\ref{eq:formal-g}), as follows:
  \begin{eqnarray}
    && \hspace{-5mm}
    \tilde{\gamma}_{n L}^2(E_{c.m.})
    \nonumber \\
    &=&
    \frac{\gamma_{n L}^2}{1-\gamma_{n L}^2 \Delta_L^\prime(E_{n L}, a_c)  \left[\, 1+Q_{n L}(E_{c.m.}, a_c)\,\right]},
    \label{eq:g-ecm}
    \\
    && \nonumber \\
    && \hspace{-5mm}    
    Q_{n L}(E_{c.m.}, a_c)
    \nonumber \\
    &=&
    \frac{1}{\Delta_L^\prime(E_{n L}, a_c)}
    \left(\frac{\Delta_L(E_{c.m.}, a_c)-\bar{\Delta}_{n L}(E_{c.m.}, a_c)}{E_{c.m.}-E_{n L}}\right)
    \nonumber\\
    &=&
    (E_{c.m.}-E_{n L})\,q_0 \left(\,1+q_1 E_{c.m.}+q_2 E_{c.m.}^2+ \dots \right),
    \label{eq:q-ecm}    
  \end{eqnarray}
where $\bar{\Delta}_{n L}$ denotes the shift function in the linear approximation.
$q_m$ ($m=0,1,2,\dots$) are the coefficients of the expansion.
If the reduced width is narrow, the formal reduced width is found to be almost identical to the observed reduced width, $\tilde{\gamma}^2_{n L} \approx \gamma^2_{n L}$.
In contrast, the formal resonance parameters are varied on energies with $Q_{n L}(E_{c.m.}, a_c)$ if the observed reduced width is large.
This means that Eq.~(\ref{eq:g-ecm}) bears the deviation from the assumed compound nuclei in \cite{Lan58}.
The higher-order correction term of Eq.~(\ref{eq:q-ecm}) is shown in Fig.~\ref{fig:coulf}(b).
The solid curves are the calculated values for the 1$^-_1$ and 1$^-_2$ states.
$Q_{n L}=0$ is used in the linear approximation.
The linear approximation is confirmed to be available only around $E_{n L}$.
The formal energy $\tilde{E}_{n L}$ of the $n$th pole including $Q_{nL}$ is defined in
\begin{eqnarray}
  &&\hspace{-5mm}
  \tilde{E}_{n L}(E_{c.m.})
  \nonumber \\
  &=& E_{n L} + \tilde{\gamma}_{n L}^2(E_{c.m.})\Delta_L(E_{n L}, a_c) \,[\,1+d_{n L}\,],
  \label{eq:er-ecm}
\end{eqnarray}
where $d_{n L}$ is a parameter stemming from the multi-poles in the $R$-matrix.
$d_{n L}$ is adjusted self-consistently so as to satisfy the relation of
\begin{eqnarray}
  \Delta_L(E_{nL},a_c) R_L(E_{nL}) &=& 1,
  \label{eq:SR}
\end{eqnarray}
where $R_L$ is the $R$-matrix defined in Eq.~(\ref{eq:rmat}).

\section{Results}

  In this section, I illustrate an example of the reduced $E$1 $S$-factor of $^{12}$C($\alpha$,$\gamma_0$)$^{16}$O by using the $R$-matrix method.
  In addition, I show the corresponding results of the $\beta$-delayed $\alpha$-particle spectrum of $^{16}$N and the $p$-wave phase shift for $\alpha$+$^{12}$C elastic scattering.
  After discussing the example, I assess the sensitivity to $a_c$ in the $E$1 $S$-factor.

\subsection{An example of the reduced $E$1 $S$-factor}

\begin{table*}[t]
  \caption{\label{tb:obprm}
      Resonance parameters used in the $R$-matrix method for $a_c= 4.75$ fm.
      The $E_{n L}$ and $\gamma_{n L}$ are the observed resonance energy and observed reduced $\alpha$-particle width, respectively.
      The values of $E_{n 1}$ are taken from \cite{Til93}.
      $\gamma_{11}$ and $\gamma_{21}$ are derived from the potential model.
      The $\Gamma_{n L}$ and $\theta_{n L}^2$ are the $\alpha$-particle width (Eq.~(\ref{eq:ga})) and the dimensionless width (Eq.~(\ref{eq:towig})).
      The $\tilde{E}_{n L}$ and $\tilde{\gamma}_{n L}$ are the formal resonance energy and formal reduced $\alpha$-particle width, respectively.
      The $\tilde{E}_{n L}$ and $\tilde{\gamma}_{n L}$ at $E_{n L}$ are listed, and they are the same as those in the linear approximation.
      The $\alpha$+$^{12}$C threshold in $^{16}$O is located at $E_x= 7.162$ MeV.
  }
  \begin{ruledtabular}
    \begin{tabular}{ccccccc}
     $L^\pi_n$  & $E_{n L}$ (MeV)& $\gamma_{n L}$ (MeV)$^{1/2}$ & $\Gamma_{n L}$ (keV) & $\theta^2_{n L}$ &
         $\tilde{E}_{n L}$ (MeV)& $\tilde{\gamma}_{n L}$ (MeV)$^{1/2}$ \\
    \hline
     1$^-_1$ & -0.0451 &  0.345  &     & 0.128 & -0.0392  &  0.355\\
     1$^-_2$ &  2.434  &  0.850  & 432 & 0.780 & -0.715   &  1.359\\
     1$^-_3$ &  5.278  &  0.150  & 100 & 0.024 &  5.267   &  0.150\\
     1$^-_4$ &  5.928  & -0.073  &  28 & 0.006 &  5.926   & -0.073\\
    \end{tabular}
  \end{ruledtabular}
\end{table*}

\begin{figure}[th]
  \centerline{
    \includegraphics[width=0.75\linewidth]{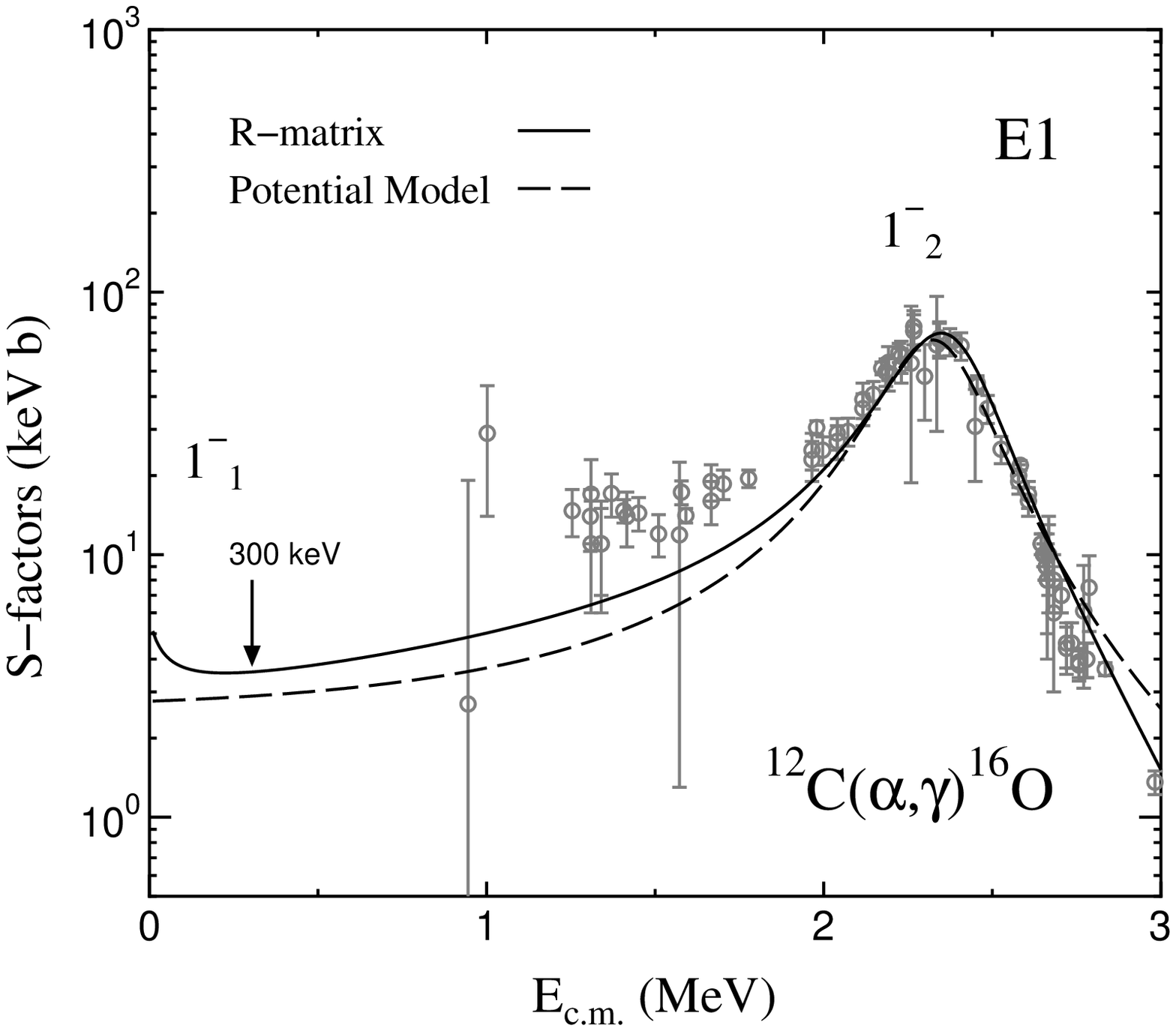}
  }
  \caption{\label{fig:s}
    Example of the reduced $E$1 $S$-factors for $^{12}$C($\alpha$,$\gamma_0$)$^{16}$O.
    The solid curve is the result obtained from the present $R$-matrix method.
    The resonance parameters are listed in Table~\ref{tb:obprm}.
    $a_c=4.75$ fm is used as the channel radius.
    The arrow indicates the energy corresponding to the most important Helium burning temperature.
    The dashed curve is the result \cite{Kat08} from the potential model.
    The experimental data are taken from \cite{Kun01,Ass06,Pla12,Oue96,Mak09}.
  }
  \vspace{6mm}
  \centerline{
    \includegraphics[width=0.75\linewidth]{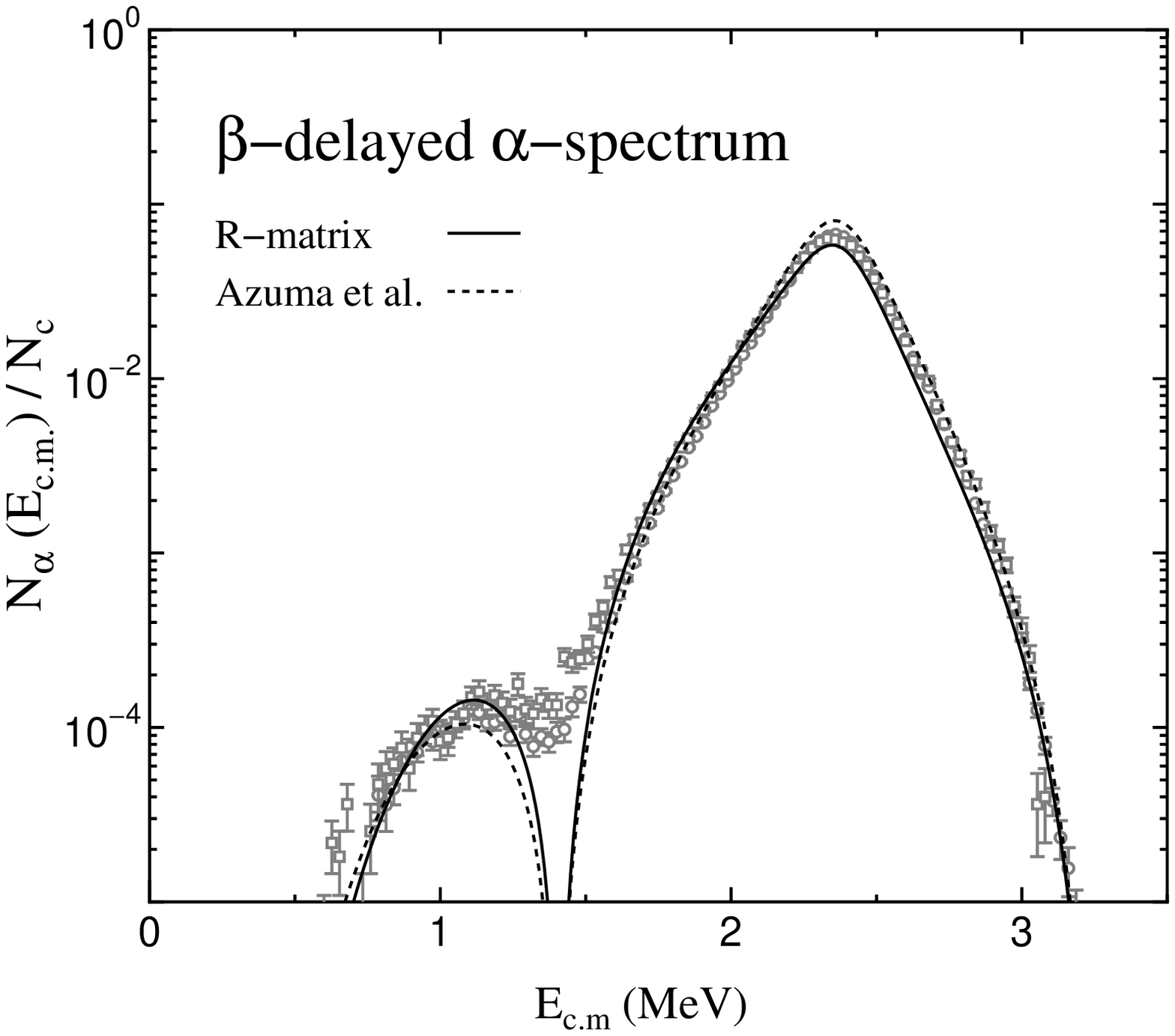}
  }
  \caption{\label{fig:b}
    $\beta$-delayed $\alpha$-particle spectrum of $^{16}$N.
    The solid curve is the result obtained from the present $R$-matrix method.
    The resonance parameters are listed in Table~\ref{tb:obprm}.
    $a_c=4.75$ fm is used as the channel radius.
    The dotted curve is the result from the $R$-matrix method for $a_c=6.5$ fm in Ref.~\cite{Azu94}.
    The experimental data are taken from ~\cite{Azu94,Tan10}.
  }
\end{figure}

  The solid curve in Fig.~\ref{fig:s} shows an example of the reduced $E$1 $S$-factor from the $R$-matrix method.
  The resonance parameters used here are listed in Table~\ref{tb:obprm}.
  $a_c=4.75$ fm is used as the channel radius.
  It should be noted that the subthreshold 1$^-_1$ state is explicitly included in the calculation.
  The arrow indicates the astrophysical energy corresponding to the most important helium burning temperature.
  From Fig.~\ref{fig:s}, I find that the $E$1 $S$-factor is not strongly enhanced at low energies, even if the subthreshold state is included.
  In this example, the $E$1 $S$-factor is approximately 3.6 keV~b at $E_{c.m.}=300$ keV, which is different from the recent estimations:
  e.g. ($100 \pm 28$) keV~b \cite{Oul12}, ($84 \pm 21$) keV~b \cite{Tan10}, ($80 \pm 18$) keV~b \cite{Nacre2}, and ($98 \pm 7$) keV~b \cite{Zhe15},
  as a foregone conclusion.
  The dashed curve is the result \cite{Kat08} from the potential model.
  The present result seems to be consistent with the potential model.
  The calculated values around $E_{c.m.}=1.5$ MeV deviate from the experimental data.
  However, the reduced $E$1 $S$-factor appears to be advocated by the recent $\gamma$-ray angular distribution of $^{12}$C($\alpha$,$\gamma_0$)$^{16}$O \cite{Kun01,Ass06,Kat08}.

\begin{figure}[t]
  \centerline{
    \includegraphics[width=0.75\linewidth]{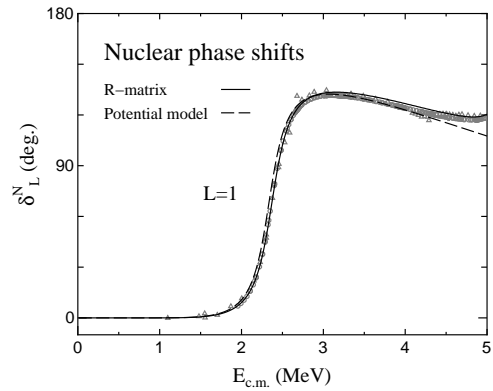}
  }
  \caption{\label{fig:ps}
    Nuclear phase shift of $L=1$ for $\alpha$+$^{12}$C elastic scattering.
    The solid curve is the result obtained from the present $R$-matrix method.
    The resonance parameters are listed in Table~\ref{tb:obprm}.
    $a_c=4.75$ fm is used as the channel radius.
    The dashed curve is the result \cite{Kat08} from the potential model.
    The experimental data are taken from \cite{Pla87,Tis09}.
  }
\end{figure}

  The corresponding $\beta$-delayed $\alpha$-particle spectrum of $^{16}$N is shown in Fig.~\ref{fig:b}.
  The solid curve is the present result from the $R$-matrix method.
  The $\beta$-feeding amplitude for the 1$^-_1$ state is obtained from the $\beta$-decay branching ratio (Eq.~(\ref{eq:b11})).
  The amplitude for the 1$^-_2$ state and background are optimized so as to fit the experimental data \cite{Azu94,Tan10}.
  The resultant values are $B_{11}/\sqrt{N_c}=1.30$, $B_{21}/\sqrt{N_c}=-0.977$, and ${\cal R}_{\beta 1}/\sqrt{N_c}=-0.276$ (MeV)$^{-1/2}$.
  The dotted curve is the result for $a_c=6.5$ fm in Ref.~\cite{Azu94}.
  The $f$-wave contribution is not included in the present article.
  This is because the predominance of $L=3$ component cannot be found in the $\alpha$+$^{12}$C continuum state at $E_{c.m.}\approx 1.3$ MeV \cite{Kat10a}.
  The present result is consistent with the published results \cite{Azu94,Tan10}.
  The nuclear phase shift of $L=1$ for elastic scattering is displayed in Fig.~\ref{fig:ps}.
  The solid curve is calculated from the sum of the $R$-matrix phase shift (Eq.~(\ref{eq:ps-r})) and hard-sphere phase shift (Eq.~(\ref{eq:ps-hs})).
  The dashed curve is the result \cite{Kat08} obtained from the potential model.
  The present result appears to be concordant with the dashed curve and the experimental phase shifts \cite{Pla87,Tis09}.

\begin{figure}[t]
  \centerline{
      \includegraphics[width=0.75\linewidth]{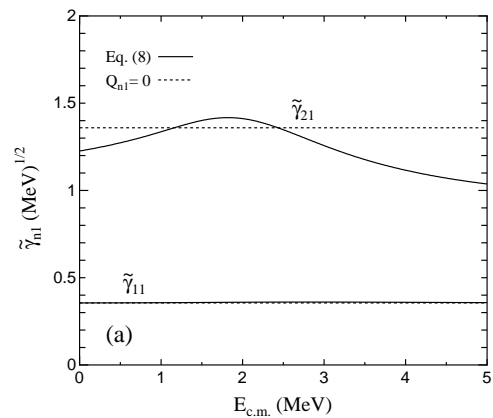} 
  }
  \caption{\label{fig:diff}
    Difference in the formal reduced widths.
    The solid curves are the formal reduced widths  $\tilde{\gamma}_{n 1}$ including the higher-order correction (Eq.~(\ref{eq:g-ecm})).
    The $\tilde{\gamma}_{2 1}$ varies with $E_{c.m.}$ because the observed reduced width is large for the 1$^-_2$ molecular state.
    The dotted lines are obtained from the linear approximation (Eq.~(\ref{eq:formal-g})).
    $a_c=4.75$ fm is used as the channel radius.
  }
\end{figure}

  The formal reduced width is shown in Fig.~\ref{fig:diff}.
  The dotted lines are obtained from the linear approximation at the resonance energies.
  The higher-order correction $Q_{n1}$ of Eq.~(\ref{eq:q-ecm}) is included in the solid curves.
  From Fig.~\ref{fig:diff}, the value of $\tilde{\gamma}_{1 1}$ is found to be identical to the dotted line of the linear approximation.
  In contrast, the value of $\tilde{\gamma}_{2 1}$ varies with $E_{c.m.}$ around the constant of the linear approximation.
  This is because the observed reduced width is large.
  So, I confirm that the linear approximation for the 1$^-_2$ state does not work well in the $R$-matrix calculation with $a_c=4.75$ fm.
  Probably, the linear approximation worked well in the previous $R$-matrix analyses because the value of $\gamma_{2 1}$ was relatively small around $a_c=6.5$ fm.
  Conversely, one might say that the channel radius should be set at $a_c \ge 5.5$ fm in order to ensure this approximation, as the approximation seems to be implemented almost implicitly in the $R$-matrix code.
  Ref.~\cite{Des} also points out that linear approximation is not valid below $a_c\approx 5$ fm.
  The $\beta$-delayed $\alpha$-particle spectrum of $^{16}$N is sensitive to the reduced widths of the 1$^-_1$ and 1$^-_2$ states.
  The reduced widths and the $E$1 $S$-factor were assessed with this sensitivity (e.g.~\cite{Azu94,Azu97,Buc96a,Buc06,Zha93a,Tan07,Tan10}).
  If the large reduced width was taken into account, the $E$1 $S$-factor might have been reduced further from the popular evaluated value.

  If the linear approximation is not valid, the formal and/or observed width depends on energy because of $Q_{nL}(E_{c.m.}, a_c)$ in Eq.~(\ref{eq:g-ecm}).
  I assume that the observed parameters are energy-independent because it is quite reasonable that the experimental nuclear structure data is energy-independent.
  So, the formal parameters are varied on energies, as shown in Fig.~\ref{fig:diff}.
  This may not appear to be allowed by the definition in \cite{Lan58}.
  However, there is no large difference from \cite{Lan58} even if the observed parameters remain energy-independent.
  In fact, the wavefunction satisfies the orthogonality in the internal region.
  Conversely, the observed parameters depend on energy, if the energy-independent formal parameters are adopted. i.e. the Breit-Wigner parameters depend on energy due to $Q_{nL}(E_{c.m.}, a_c)$.
  The derived energy-dependence seems model-dependent.

\begin{figure}[t]
  \begin{center}
    \includegraphics[width=0.75\linewidth]{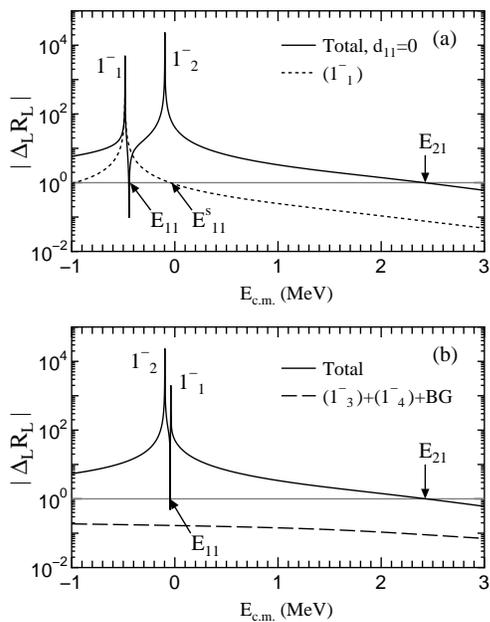}
  \end{center}
  \caption{\label{fig:SR}
    Resultant $R$-matrix multiplied by $\Delta_L(E_{c.m.})$.
    $a_c=4.75$ fm is used as the channel radius.
    The solid curve in panel (a) is obtained with $d_{11}=0$ of Eq.~(\ref{eq:er-ecm}).
    The dotted curve is the pure component of the 1$^-_1$ state.
    The observed resonance energy of the 1$^-_1$ state is shifted from $E^s_{11}$ because of the interference with the broad 1$^-_2$ resonance.
    In panel (b), the pole energy of 1$^-_1$, $\tilde{E}_{11}$, is adjusted with Eq.~(\ref{eq:SR}) to obtain the better observed resonance energy $E_{11}$.
    $d_{11}=-1.0133$ is used.
    The interference between 1$^-_1$ and 1$^-_2$ seems weak.
    The dashed curve is the sum of the components for the background, 1$^-_3$, and 1$^-_4$.
  }
\end{figure}

  The derived $R$-matrix multiplied by $\Delta_L$ is illustrated in Fig.~\ref{fig:SR}.
  The solid curve in Fig.~\ref{fig:SR}(a) is the total component obtained with $d_{nL}=0$ of Eq.~(\ref{eq:er-ecm}).
  The peak corresponds to the formal energy, and the energy position of $\Delta_L R_L=1$ corresponds to the observed energy.
  The pole of the 1$^-_2$ state is located beneath the $\alpha$-particle threshold, because the large energy shift is generated by the large reduced width of the $\alpha$+$^{12}$C molecular state.
  The dotted curve represents the pure single pole component of the 1$^-_1$ state.
  The energy $E^s_{11}$ in Fig.~\ref{fig:SR}(a) indicates the observed resonance energy of 1$^-_1$ in the single pole approximation.
  However, this energy position is shifted lower by the interference with the broad 1$^-_2$ resonance.
  Consequently, the calculated $E_{11}$ from the whole components of $R$-matrix is different from $E^s_{11}$.
  So, $\tilde{E}_{11}$ is re-defined with Eqs.~(\ref{eq:er-ecm}) and (\ref{eq:SR}), so as to make the appropriate energy of $E_{11}$.
  The resulting $R$-matrix is shown by the solid curve in Fig.~\ref{fig:SR}(b).
  $d_{11}=-1.0133$ is used, and the $\tilde{\gamma}_{n L}$ and $\tilde{E}_{n L}$ at $E_{n L}$ are listed in Table~\ref{tb:obprm}.
  The narrow peak of 1$^-_1$ is on the broad resonance of 1$^-_2$.
  In addition, the pole of 1$^-_1$ is located in the vicinity of 1$^-_2$ below $E_{c.m.}= 0$.
  The interference between two poles is weak.
  The proximity of two poles seems to suppress the interference.
  The 1$^-_2$ resonance dominates the $R$-matrix below the barrier.
  This weak interference reduces the contribution from the subthreshold 1$^-_1$ state.

  The suppression of the interference appears to originate from the difference in strength of the $R$-matrix component between 1$^-_1$ and 1$^-_2$.
  The 1$^-_2$ state holds the dominant component at the energies except for $\tilde{E}_{11}$ of 1$^-_1$, even when they are close together.
  However, the contribution from 1$^-_2$ becomes weak around $\tilde{E}_{11}$, when 1$^-_2$ stays away from 1$^-_1$.
  So, the 1$^-_1$ state makes the prominent contribution, together with the interference with the tail of 1$^-_2$.

  The channel radius of the present example is shorter than $a_c= 6.5$ fm used widely in the $R$-matrix analyses.
  I, however, think that $a_c\approx 4.2$ -- 5.2 fm is acceptable, because the classical turning point after penetrating the barrier is approximately $r=5$ fm \cite{Kat10b} and because the rough estimation of the contact distance is $\langle r^2\rangle^{1/2}_{\rm 12C} +\langle r^2\rangle^{1/2}_\alpha\approx 4.2$ fm from the root-mean-square radius of nuclei \cite{Dev87}.
  The present $R$-matrix method has the boundary condition of $\tilde{b}_c=0$ (Eq.~(\ref{eq:bc})) and $b_c \equiv \varphi^\prime(a_c)/\varphi(a_c) \approx 0$ (Eq.~(\ref{eq:bnl})).
  So, $a_c$ is approximately equivalent to the position of the first peak of the probability after the barrier penetration.
  The transparency of the $\alpha$+$^{12}$C system, i.e. the weak interference between $\alpha$+$^{12}$C and others, is found to be expressed as the reaction with the shrinking interaction region.
  
\begin{figure}[t]
  \centerline{
    \includegraphics[width=0.75\linewidth]{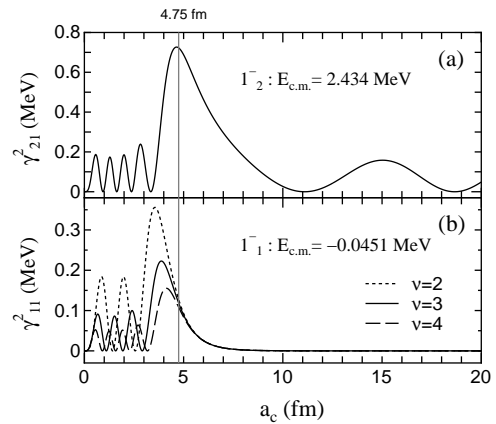}
  }
  \caption{\label{fig:wf}
    Reduced widths $\gamma^2_{n L}$ for 1$^-_2$ and 1$^-_1$, as a function of the channel radius $a_c$.
    The thin line represents the adapted channel radius $a_c=4.75$ fm.
    (a) $\gamma_{21}^2$ is obtained from the wavefunction of potential scattering at $E_{c.m.}=2.434$ MeV \cite{Kat10b}.
    (b) $\gamma_{11}^2$ is calculated from the bound state wavefunction with the different radial node $\nu$ \cite{Kat08,Kat10b}: $\nu=2$ (dotted curve), $\nu=3$ (solid curve), and $\nu=4$ (dashed curve).
  }
\end{figure}

  To clarify the boundary condition, the reduced widths $\gamma^2_{n L}$ for 1$^-_2$ and 1$^-_1$ are illustrated in Fig.~\ref{fig:wf}, as a function of $a_c$.
  The thin line represents the position of the adapted channel radius $a_c=4.75$ fm.
  $\gamma_{21}^2$ is obtained from the wavefunction of potential scattering at $E_{c.m.}=2.434$ MeV \cite{Kat10b}.
  In Fig.~\ref{fig:wf}(a), the maximum peak of $\gamma_{21}^2$ corresponds to the adopted channel radius.
  The $\alpha$-particle width is obtained from Eq.~(\ref{eq:ga}) with Fig.~\ref{fig:coulf}(a) and Fig.~\ref{fig:wf}(a); $\Gamma_\alpha=432$ keV at $a_c=4.75$ fm.
  For the 1$^-_1$ state, $\gamma_{11}^2$ is presumed from the bound state wavefunction with the radial node $\nu$ \cite{Kat08}.
  Figure~\ref{fig:wf}(b) shows the values of $\gamma_{11}^2$ with $\nu=2$ (dotted curve), $\nu=3$ (solid curve), and $\nu=4$ (dashed curve).
  Three curves are almost identical for $a_c \ge 4.75$ fm.
  The maximum value of $\gamma_{11}^2$ may be given around $a_c=4.75$ fm, because $\gamma_{11}^2$ is independent of $\nu$.
  The realistic value may be small in the internal region if the 1$^-_1$ state is not described with the $\alpha$+$^{12}$C configuration.
  Suppose $\gamma_{11}^2$ is the maximum at $a_c=4.75$ fm, the 1$^-_1$ state also seems to satisfy $b_c\approx 0$ at this radius.

  The small $E$1 $S$-factor at low energies could be found in literature.
  In Figs.~16 and 18 of Ref.~\cite{Azu94}, the $\chi^2$ minimum can be found at the small $E$1 $S$-factor.
  The photo-nuclear reaction of $^{16}$O($\gamma$,$\alpha$)$^{12}$C also expects the small $E$1 values \cite{Gai14}.
  The small $E$1 $S$-factor with the deep dip below the barrier can be made by the strong destructive interference between the 1$^-_1$ and 1$^-_2$ states (e.g.~\cite{Oue96,Gia01,Ham05a}).
  The better reproduction of the $\beta$-delayed $\alpha$-particle spectrum data and the reduced $E$1 $S$-factor have been reported to be obtained with the complex $\beta$-decay feeding amplitude \cite{Hal97}.
  The present $R$-matrix calculation might be one of the relatives of the previous analyses mentioned here.
  It is, however, noted that $\gamma_{11}$ and $\gamma_{21}$ in the present article are derived from the $\alpha$+$^{12}$C potential model.
  In addition, I include the higher-order correction to the linear approximation.
  So, I reckon that the small $E$1 $S$-factor can be steadied by Eq.~(\ref{eq:q-ecm}).
  The $R$-matrix method does not depend on the procedure for generating the internal wavefunction.
  If the similar boundary is obtained from other theoretical models, the corresponding $E$1 $S$-factor would not be enhanced by the subthreshold state.

\subsection{Sensitivity to the channel radius in the $E$1 $S$-factor}

\begin{figure}[t]
  \centerline{
    \includegraphics[width=0.75\linewidth]{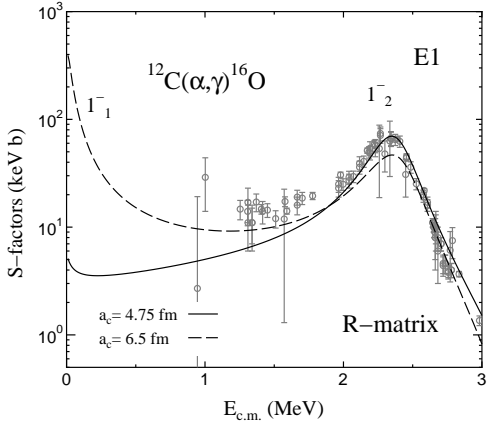}
  }
  \caption{\label{fig:dif-on-ac}
    Sensitivity to the channel radius in the $E$1 $S$-factor for $^{12}$C($\alpha$,$\gamma_0$)$^{16}$O.
    The solid and dashed curves are obtained from $a_c= 4.75$ fm (Table~\ref{tb:obprm}) and $a_c=6.5$ fm (Table~\ref{tb:obprm-b}), respectively.
    The experimental data are taken from \cite{Kun01,Ass06,Pla12,Oue96,Mak09}.
  }
\end{figure}

\begin{figure}[t]
  \begin{center}
    \begin{tabular}{c}
      \includegraphics[width=0.65\linewidth]{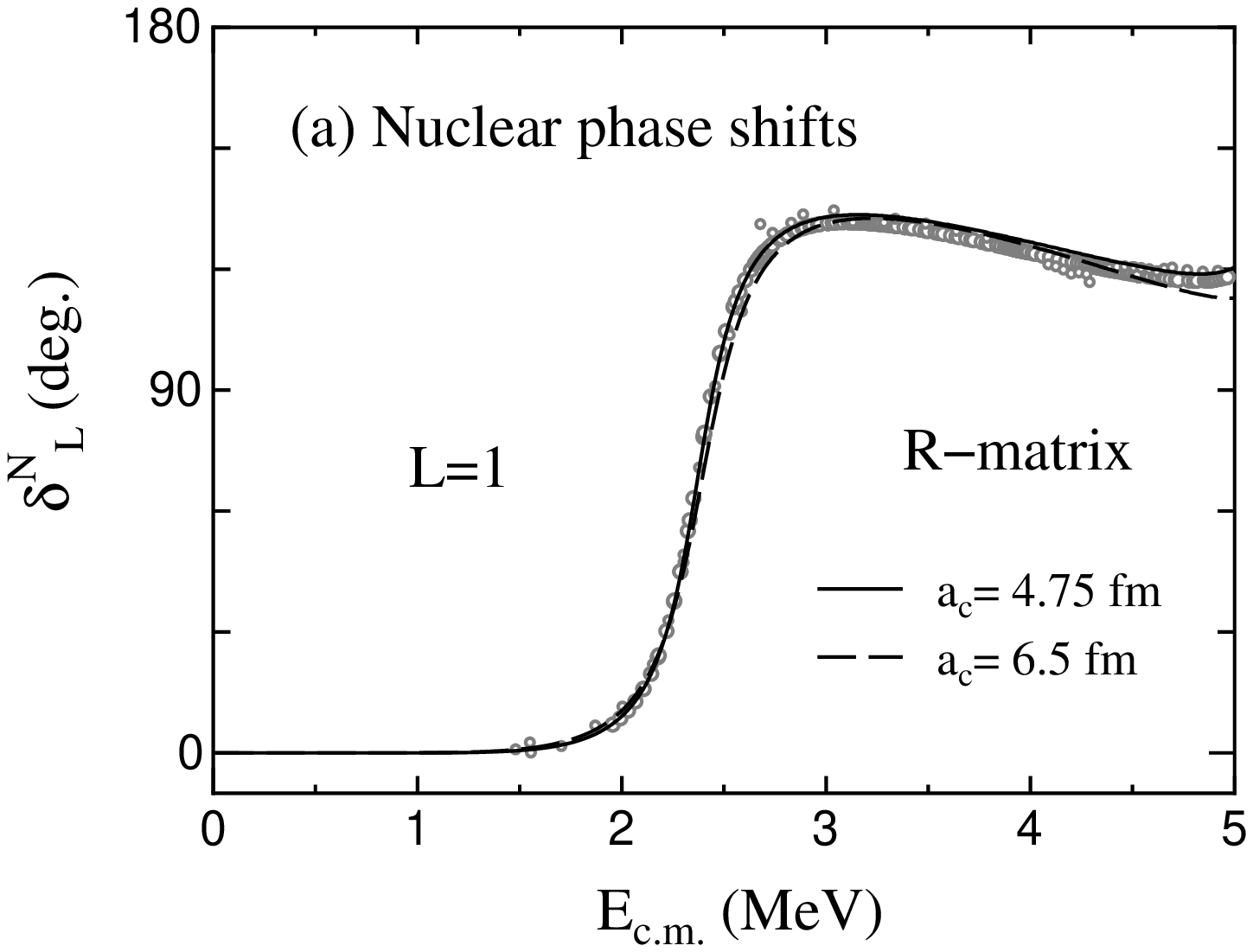} \\
      \includegraphics[width=0.65\linewidth]{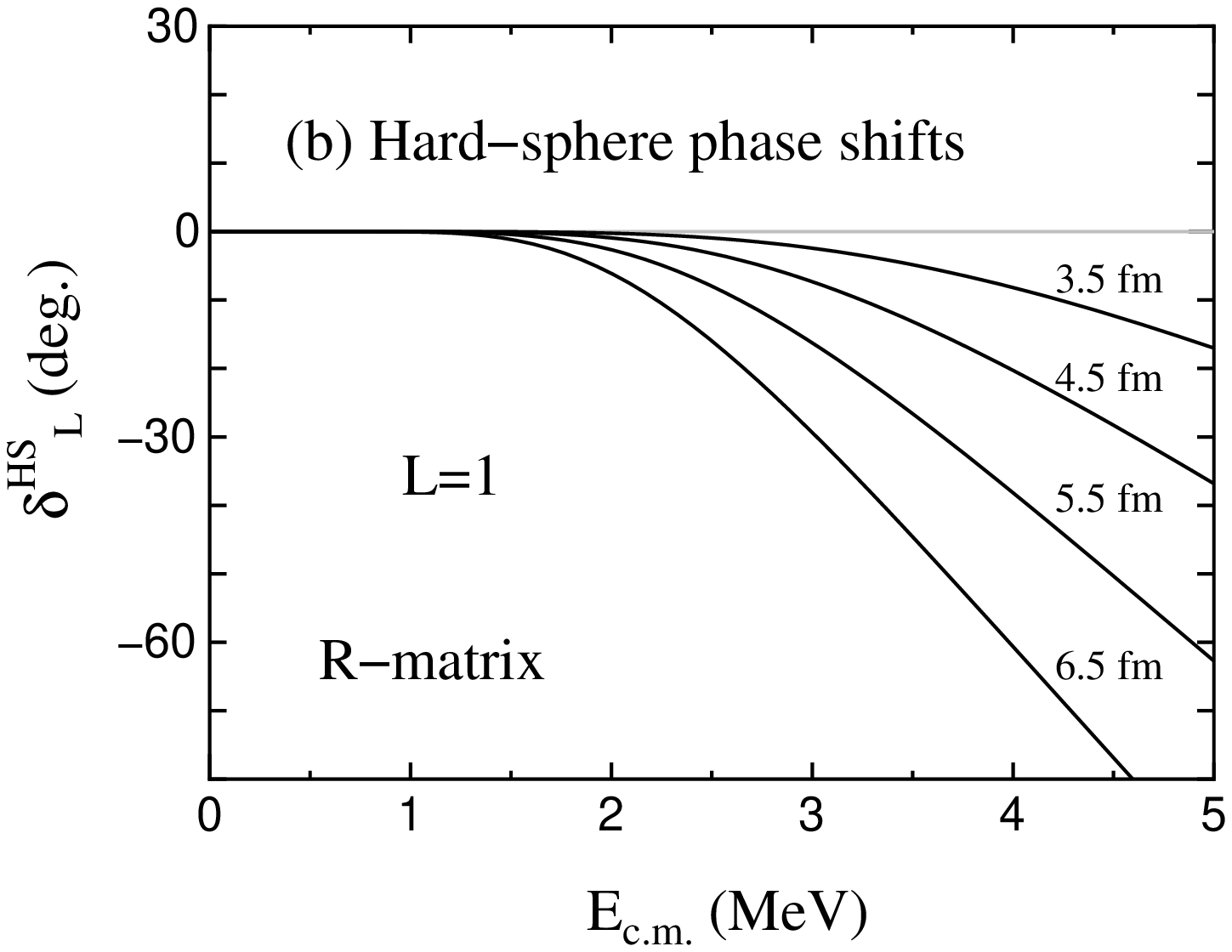} \\
    \end{tabular}
  \end{center}
  \caption{\label{fig:difps}
    (a) Nuclear phase shift of $L=1$ for $\alpha$+$^{12}$C elastic scattering.
    The solid and dashed curves are the results obtained from the $R$-matrix method with $a_c=4.75$ fm and 6.5 fm.
    The resonance parameters are listed in Tables~\ref{tb:obprm} and \ref{tb:obprm-b}.
    The experimental data are taken from \cite{Pla87,Tis09}.
    (b) Hard-sphere phase shift of $L=1$ for $\alpha$+$^{12}$C elastic scattering.
    The solid curves are calculated from the channel radius $a_c= 3.5$, 4.5, 5.5, and 6.5 fm.
  }
\end{figure}

  Figure~\ref{fig:dif-on-ac} shows the sensitivity to the channel radius in the $E$1 $S$-factor for $^{12}$C($\alpha$,$\gamma_0$)$^{16}$O.
  The solid and dashed curves are the calculated results with $a_c=4.75$ fm (Table~\ref{tb:obprm}) and $a_c=6.5$ fm (Table~\ref{tb:obprm-b}), respectively.
  The solid curve is the same as that in Fig.~\ref{fig:s}, and the $E$1 $S$-factor is not enhanced at low energies.
  However, I find from this figure that it is enhanced by the subthreshold 1$^-_1$ state if $a_c=6.5$ fm is used.
  The corresponding nuclear phase shift of $L=1$ for elastic scattering is shown in Fig.~\ref{fig:difps}(a).
  The experimental phase shift data appear to be reproduced within the same quality.

  The channel radius manipulates the enhancement of the $E$1 $S$-factor at low energies.
  The large channel radius expands the strong interacting region, along with the high penetrability of the Coulomb barrier, and it makes the process more reactive.
  The contribution from the state below the barrier is consequently magnified, even though the dimensionless width becomes small.
  The use of the large channel radius is equivalent to the reduction of the barrier, and it apparently makes the strong interference between the states.

  One may think that the channel radius $a_c$ is the parameter for convenience of calculation, and that the calculated physical observable should be independent of the artificial parameter.
  Surely, $a_c$ is the adjustable free parameter, but at the same time it defines the radius of the strong absorptive region or strongly interacting region.
  After the fit to the experimental data, the calculated result is insensitive to the small variation of $a_c$ involving other buffer parameters.
  Furthermore, the consistent description of multiple observables, e.g. the phase shifts and cross sections, appears to impose a constraint on $a_c$.

  The background resonance makes the theoretical patchwork to the assumed interacting region with sharp-cut edge.
  The phase shift of hard-sphere scattering appears in the negative angles, $\delta_L^{\rm HS} < 0$, as shown in Fig.~\ref{fig:difps}(b).
  In addition, if the channel radius is large, the absolute value of $\delta_L^{\rm HS}$ becomes large for $E_{c.m.} \ge 2$ MeV.
  Consequently, the broad background resonance is required to cancel out the hard-sphere phase shift (e.g.~\cite{Ang00}).
  If the background resonance is adjusted well, the $R$-matrix calculation with any choice of $a_c$ may reproduce the phase shift data in the same quality.
  It is, however, noted that the calculation exceeds the Wigner sum rule limit \cite{Wig52} substantially because of the hypothetical background state.
  To avoid the confusion, I adopted the energy independent background in Eq.~(\ref{eq:rmat}).
  The background is ${\cal R}_{\alpha 1}= 0.0432$ for $a_c=4.75$ fm and ${\cal R}_{\alpha 1}= 0.2065$ for $a_c=6.5$ fm.
  For $a_c=4.75$ fm in $E_{c.m.}<3$ MeV, the contribution from the background is small, namely, the absolute value of the sum of the $R$-matrix for the background, 1$^-_3$, and 1$^-_4$ is less than 10\% of the magnitude for 1$^-_1$ and 1$^-_2$. (dashed curve in Fig.~\ref{fig:SR}(b))
  On the other hand, the background for $a_c=6.5$ fm affects the result below the barrier, considerably.

\begin{table*}[t]
  \caption{\label{tb:obprm-b}
    Same as Table~\ref{tb:obprm}, but for $a_c= 6.5$ fm.
  }
  \begin{ruledtabular}
    \begin{tabular}{ccccccc}
     $L^\pi_n$  & $E_{n L}$ (MeV)& $\gamma_{n L}$ (MeV)$^{1/2}$ & $\Gamma_{n L}$ (keV) & $\theta^2_{n L}$ &
         $\tilde{E}_{n L}$ (MeV)& $\tilde{\gamma}_{n L}$ (MeV)$^{1/2}$ \\
    \hline
     1$^-_1$ & -0.0451 & -0.100  &     & 0.020 & -0.0603 & -0.100\\
     1$^-_2$ &  2.423  &  0.533  & 570 & 0.574 &  2.061  &  0.617\\
     1$^-_3$ &  5.278  &  0.105  &  85 & 0.022 &  5.275  &  0.105\\
     1$^-_4$ &  5.928  & -0.057  &  28 & 0.007 &  5.927  & -0.057\\
    \end{tabular}
  \end{ruledtabular}
\end{table*}

  In Table \ref{tb:obprm}, the observed $\gamma_{21}$ for the 1$^-_2$ state is the largest of the considered states.
  This is because the 1$^-_2$ state is the member of the $\alpha$+$^{12}$C molecular bands.
  The present $R$-matrix reproduces the observed width of 1$^-_2$, $\Gamma_{21}=(420\pm20)$ keV \cite{Til93}.
  This means that the calculated 1$^-_2$ state has the appropriate $\alpha$-decay property.
  In contrast, $\Gamma_{21}$ from the present model with $a_c=6.5$ fm is $\Gamma_{21}=570$ keV because the reduced width is large for the molecular state. (Table \ref{tb:obprm-b})
  Therefore, the present result for $a_c=6.5$ fm is discarded.
  The previous $R$-matrix analyses gave the slightly narrow $\alpha$-particle width of 1$^-_2$:
  e.g. $\Gamma_{21}= 359$ keV \cite{Azu94}, $\Gamma_{21}= 388$ keV \cite{Oul12}, and $\Gamma_{21}= 322$ keV \cite{Zhe15}.
  These are obtained from $a_c=6.5$ fm.
  The $\alpha$-decay property of 1$^-_2$ does not seem to be reproduced well in \cite{Azu94,Oul12,Zhe15}.
  This may have been caused by the implicit assumption that the $^{12}$C($\alpha$,$\gamma$)$^{16}$O reaction would happen in compound nucleus reactions for low energies.

  The $\theta_{11}^2$ is derived from ANC of the 1$^-_1$ state corresponding to the $\alpha$-particle spectroscopic factor, $S_\alpha\approx 0.3$ \cite{Kat08}.
  $\theta_{11}^2= 0.128$ is obtained for $a_c=4.75$ fm.
  This value seems quite large.
  $\theta_{11}^2=0.020$ for $a_c=6.5$ fm in Table \ref{tb:obprm-b} appears to be comparable with e.g.
  $\theta_{11}^2\approx 0.016$ \cite{Oul12}, 0.0096--0.0166 \cite{Bel07}, 0.017 \cite{Bru99}, and 0.013 \cite{Azu94,Tan10,Zhe15}.
  The surface probability of $\alpha$-particle obviously depends on the channel radius.
  Not only the $\alpha$+$^{12}$C configuration in the 1$^-_1$ state but also the channel radius is significant in the determination of the $E$1 $S$-factor at $E_{c.m.}=300$ keV.
  The position of ``surface'' influences the destinies of the $R$-matrix calculation.
  The ANC can be determined from the indirect measurements.
  I, however, reckon that their analyses created the strong $E1$ enhancement of the $S$-factor when the $R$-matrix code was invoked.

  In the present article, the reduced $E$1 $S$-factor at low energies is exemplified by the small interacting region along with the low penetrability, in addition to the well-developed 1$^-_2$ molecular resonance.
  Judging from the decay property and the equivalence of the potential model, I consider that the reduced $E$1 $S$-factor would rather be preferable than the strong $E$1 enhancement.

\section{Summary}
  
  I have exemplified the low-energy $E$1 $S$-factor of $^{12}$C($\alpha$,$\gamma_0$)$^{16}$O with the $R$-matrix theory.
  The reduced $\alpha$-particle widths of the 1$^-_1$ and 1$^-_2$ states are extracted from the potential model.
  The formal parameters are examined with the higher-order correction to the shift function.
  The correction enables me to use the small channel radius usually discarded.
  I have also assessed the sensitivity to the channel radius $a_c$.

  As an example, I have illustrated the reduced $E$1 $S$-factor with $a_c= 4.75$ fm.
  The corresponding $\beta$-delayed $\alpha$-particle spectrum of $^{16}$N and the $p$-wave phase shift for $\alpha$+$^{12}$C elastic scattering are consistent with the previous studies.
  The adopted channel radius is shorter than that used in the previous $R$-matrix analyses.
  However, I have found that the transparency of the $\alpha$+$^{12}$C system, i.e. the weak interference between $\alpha$+$^{12}$C and others, is expressed as the reaction with the shrinking interaction region.
  The energy shift of the pole for 1$^-_2$ is large because the reduced width is large.
  The pole of 1$^-_2$ is located in the vicinity of the subthreshold 1$^-_1$ state.
  The proximity of two poles suppresses the coupling between the states.
  Under the circumstance, the low-energy $E$1 $S$-factor is not strongly enhanced even if the 1$^-_1$ state has the component of the $\alpha$+$^{12}$C configuration corresponding to $C^2=5.0\times10^{28}$ fm$^{-1}$.

  The channel radius controls the enhancement of the $E$1 $S$-factor at low energies.
  If $a_c = 6.5$ fm is used, the $E$1 $S$-factor is enhanced by the subthreshold state.
  The large $a_c$ leads to the high penetrability of the Coulomb barrier, and it facilitates the interference between the 1$^-_1$ and 1$^-_2$ states.
  However, the $\alpha$-decay property of 1$^-_2$ is not reproduced by the model because of the large reduced width.

  The reduced $E$1 $S$-factor in the present article is consistent with the previous result \cite{Kat08} from the potential model and the experimental decay property.
  In addition, the reduction of the $E$1 transition appears to be consistent with the recent experimental results \cite{Ass06,Kun01} of the 90$^\circ$ minimum $\gamma$-ray angular distribution around $E_{c.m.}= 1.3$ MeV \cite{Kat08}.
  I, therefore, consider that the reduced $E$1 $S$-factor would rather be preferable than the $E$1 enhancement associated with the strong coupling mechanism.
  And I think that the reaction rates of $^{12}$C($\alpha$,$\gamma$)$^{16}$O are determined by the direct-capture component \cite{Kat12,Kat15}.
  In the future, the reduced $E$1 $S$-factor and reaction mechanism will be investigated in the photo-disintegration of $^{16}$O \cite{Kat14c,Kat16}, in which the cross section is expected to be measured more accurately.

\begin{acknowledgments}
I am grateful to Professor S.~Kubono for his encouragement.
I thank anonymous referees for the valuable comments on the previous version of my manuscript.
I also thank M.~Arnould, A.~Jorissen, K.~Takahashi, and H.~Utsunomiya for their hospitality during my stay at Universit\'e Libre de Bruxelles, and Y.~Ohnita and Y.~Sakuragi for their hospitality at Osaka City University.
\end{acknowledgments}

\appendix
\section{$R$-matrix method}

The $R$-matrix theory is the powerful tool to evaluate the low-energy nuclear reactions \cite{Des,Ang00,Des04,Tho09,Hum91,Lan58,Tho51}.
The conventional $R$-matrix method gives the experimental quantities by adjusting the boundary condition through the resonance parameters, without calculating the wavefunction of compound nucleus numerically.
From the importance of the nuclear surface, one may recall the compound nuclear reaction with strong coupling.
However, it also describes the single-particle motion in the low-energy reaction.
Recently, the $R$-matrix method has been used extensively to obtain the solution of low-energy nuclear reactions by the spherical interacting region with sharp-cut edge.
In the conventional method, the $R$-matrix is generated by a bunch of the resonance parameters.
In Appendix A, I describe the basic formula for the $R$-matrix theory used in the present article.

\subsection{$\alpha$+$^{12}$C elastic scattering}

The radial wavefunction of the relative motion between $\alpha$+$^{12}$C is divided into two regions at a channel radius $a_c$.
The $a_c$ is defined at the distance, where $\alpha$ and $^{12}$C are well-distinguished.
The compound nucleus is formed in the spherical region of $r < a_c$.
The wavefunction for $E_{c.m.}> 0$ in the external region ($r > a_c$) is defined as
  \begin{eqnarray}
    && \hspace{-5mm}
    \chi_L^{\rm Ext}(k_i,r)
    \nonumber \\
    &=&
    \frac{i}{2}\sqrt{\frac{2}{\pi\hbar v}} \left[\,I_L(k_i r)-U_L O_L(k_i r)\,\right],
    \label{eq:chi_out}
    \\
    &=&
    e^{i(\delta^C_L+\delta^N_L)} \sqrt{\frac{2}{\pi\hbar v}} \left[\,F_L(k_i r) \cos\delta^N_L + G_L(k_i r)\sin\delta^N_L \,\right],
    \label{eq:chi_out_re}
    \nonumber \\
  \end{eqnarray}
where $I_L(k_i r)$ and $O_L(k_i r)$ are the incoming and outgoing Coulomb wave functions, $I_L(k_i r) = O_L^\ast(k_i r) = \exp(i\delta^{\rm C}_L)[G_L(k_i r) - i F_L(k_i r)]$.
$F_L(k_i r)$ and $G_L(k_i r)$ are the regular and irregular Coulomb wave functions, respectively.
$\delta^{\rm C}_L$ and $\delta^{\rm N}_L$ are the Coulomb and nuclear phase shifts.
$L$ is the angular momentum of the relative motion between two nuclei.
$U_L$ denotes the collision matrix.
$k_i$ is the wave number, $k_i = \sqrt{2\mu E_{c.m.}/\hbar^2}$;
$\mu$ is the reduced mass.
$v$ is the velocity of the relative motion between $\alpha$ and $^{12}$C nuclei.
The wave function in the internal region ($r \le a_c$) is expanded by arbitrary orthogonal functions $\tilde{\varphi}_{n L}(r)$.
  \begin{eqnarray}
    \chi_L^{\rm Int}(r)
    &=& 
    \mathop{\sum}_n A_{n L}\tilde{\varphi}_{n L}(r),
    \label{eq:chi_in}
  \end{eqnarray}
where $A_{n L}$ is the coefficient of the expansion.

The collision matrix $U_L$ in Eq.~(\ref{eq:chi_out}) is given by,
  \begin{eqnarray}
    U_L&=&
    \frac{I_L(k_i a_c)}{O_L(k_i a_c)}
    \cdot \frac{1-[L_L^\ast(k_i a_c)-a_c \tilde{b}_c] R_L}{1-[L_L(k_i a_c)-a_c \tilde{b}_c]R_L},
    \label{eq:umat}
  \end{eqnarray}
  where $\tilde{b}_c$ is the logarithmic derivative of the internal wavefunction at $a_c$, defined in Appendix A.5.
  $L_L(k_i a_c)$ is the logarithmic derivative of the outgoing Coulomb function, and it is defined, as follows:
  \begin{eqnarray}
    L_L(k_i a_c)
    &=& a_c\frac{d}{dr} \ln O_L(k_i r)\bigr|_{r=a_c}
    \nonumber \\
    &=& \Delta_L(E_{c.m.}, a_c)+iP_L(E_{c.m.}, a_c).
  \end{eqnarray}
The real and imaginary parts of $L_L(k_i a_c)$ are the shift function $\Delta_L$ and the penetration factor $P_L$, respectively.
They are defined by
  \begin{eqnarray}
    \Delta_L(E_{c.m.}, a_c)&=& P_L(E_{c.m.}, a_c)
          [G_L(k_i a_c)G_L^\prime(k_i a_c)
            \nonumber \\
            &+&F_L(k_i a_c)F_L^\prime(k_i a_c)],
    \label{eq:shift}
    \\
    P_L(E_{c.m.}, a_c)&=&\frac{k_i a_c}{G_L^2(k_i a_c)+F_L^2(k_i a_c)}.
    \label{eq:pene}
  \end{eqnarray}
In Eq.~(\ref{eq:umat}), $R_L$ denotes $R$-matrix for elastic scattering, defined as the inverse logarithmic derivative of the internal wave function at $r=a_c$. It is expressed as
  \begin{eqnarray}
    R_L(E_{c.m.})
    &=&
    \mathop{\sum}_{n} \frac{\tilde{\gamma}_{n L}^2}{\tilde{E}_{n L}-E_{c.m.}} + {\cal R}_{\alpha L},
    \label{eq:rmat}
  \end{eqnarray}
where $\tilde{E}_{n L}$ and $\tilde{\gamma}_{n L}$ are the formal resonance energy and the formal reduced width of the $n$th resonance.
${\cal R}_{\alpha L}$ is the energy-independent background.
The contributions from poles at high excitation energies are included in the non-resonant contribution.
The collision matrix is alternatively expressed as
  \begin{eqnarray}
    U_L &=& e^{2i(\delta_L^{\rm C}+\delta^{\rm HS}_L+\delta^{\rm R}_L)},
  \end{eqnarray}
where $\delta^{\rm HS}_L$ and $\delta^{\rm R}_L$ are the hard-sphere phase shift and the $R$-matrix phase shift, respectively.
They are given in
  \begin{eqnarray}
    \delta^{\rm HS}_L &=& -\arctan \frac{F_L(k_i a_c)}{G_L(k_i a_c)},
    \label{eq:ps-hs}
    \\
    \delta^{\rm R}_L &=& \arctan \frac{P_L(E_{c.m.}, a_c)R_L(E_{c.m.})}{1-[\Delta_L(E_{c.m.}, a_c)-a_c \tilde{b}_c] R_L(E_{c.m.})}.
    \label{eq:ps-r}
    \nonumber \\
  \end{eqnarray}
The hard-sphere scattering comes from the interacting region defined by $a_c$.
So, the nuclear phase shift for elastic scattering is defined by $\delta^{\rm N}_L\equiv \delta^{\rm HS}_L+\delta^{\rm R}_L$.

The coefficient $A_{n L}$ is obtained from the continuities of the wave function at $r = a_c$.
Using Eqs.~(\ref{eq:chi_out}) and (\ref{eq:chi_in}), $A_{n L}$ is deduced in the following form,
  \begin{eqnarray}
    A_{n L} &=& 
    \frac{ e^{i(\delta_L^{\rm C}+\delta^{\rm N}_L)}}{\sqrt{2\pi}}
    \frac{ \sqrt{2P_L(E_{c.m.}, a_c)}}{|\,1 -[L_L(k_i a_c)-a_c \tilde{b}_c] R_L(E_{c.m.})\,|}
    \nonumber \\
    &\cdot& \frac{\tilde{\gamma}_{n L}}{\tilde{E}_{n L}-E_{c.m.}}.
  \end{eqnarray}

\subsection{$^{12}$C($\alpha$,$\gamma_0$)$^{16}$O reaction}

The radiative capture cross sections for $^{12}$C($\alpha$,$\gamma_0$)$^{16}$O are given by,
  \begin{eqnarray}
    \sigma^{E\lambda} (E_{c.m.})
    &=& \frac{\pi}{k_i^2} (2L+1) \left|\,T^{E\lambda}(E_{c.m.}) \,\right|^2,
    \label{eq:cap}
  \end{eqnarray}
where $T^{E\lambda}$ is the transition amplitude of the multipolarity $\lambda$.
Note that $\lambda=L$ is found in $^{12}$C($\alpha$,$\gamma_0$)$^{16}$O.
From the division of the radial integrals, $T^{EL}$ is separated into two parts,
  \begin{eqnarray}
    T^{EL}(E_{c.m.}) &=& T^{EL}_{\rm Int}(E_{c.m.}) + T^{EL}_{\rm Ext}(E_{c.m.}).
  \end{eqnarray}
The external term for $E$1 transition is vanished by the iso-spin selection rule.
The internal part is given by
  \begin{eqnarray}
    T^{EL}_{\rm Int}(E_{c.m.}) &=&
    \frac{e^{i(\delta_L^{\rm C}+\delta_L^{\rm N})} \sqrt{2P_L(E_{c.m.}, a_c)}}{|\,1 -[L_L(k_i a_c)-a_c \tilde{b}_c] R_L(E_{c.m.})\,|}
    \nonumber \\
    &\cdot&
    \mathop{\sum}_n \frac{ \tilde{\gamma}_{n L} \sqrt{\tilde{\Gamma}^{EL}_{\gamma n}}}{\tilde{E}_{n L}-E_{c.m.}}
    \left(\!\frac{E_{c.m.}\!-\!E_g}{E_{n L}\!-\!E_g}\!\right)^{\frac{2L+1}{2}},
    \nonumber \\
  \end{eqnarray}
where $\tilde{\Gamma}^{EL}_{\gamma n}$ denotes the formal $\gamma$-width.
$E_g$ is the ground-state energy, $E_g = -7.162$ MeV.
The formal $\gamma$-width is assumed to be given from the observed $\gamma$-width $\Gamma^{EL}_{\gamma n}$, as follows:
  \begin{eqnarray}
     \tilde{\Gamma}_{\gamma n}^{EL}
     &=&
     \Gamma_{\gamma n}^{EL} [\,1+\tilde{\gamma}_{n L}^2 \Delta_L^\prime(E_{nL}, a_c)\,],
     \label{eq:formal-gm}
  \end{eqnarray}
The observed $\gamma$-widths are taken from \cite{Til93,Oul12}:
$\Gamma_{\gamma 1}^{E1} = 5.5\times10^{-2}$ eV, $\Gamma_{\gamma 2}^{E1} = 1.56\times10^{-2}$ eV, $\Gamma_{\gamma 3}^{E1} = 12$ eV, and $\Gamma_{\gamma 4}^{E1} = 32$ eV.

The astrophysical $S$-factor is used, instead of the capture cross section, to compensate for the rapid energy variation below the barrier.
The $S$-factor is defined as
  \begin{eqnarray}
    {\cal S}^{EL}(E_{c.m.}) &=& E_{c.m.} \exp(2\pi\eta) \,\sigma^{EL}(E_{c.m.}),
  \end{eqnarray}
where $\eta$ is the Sommerfeld parameter, $\eta=12e^2/(\hbar v)$.

\subsection{$\beta$-delayed $\alpha$-particle spectrum of $^{16}$N}

The $\beta$-delayed $\alpha$-particle spectrum of $^{16}$N is given only from the internal region, as follows:
  \begin{eqnarray}
    && \hspace{-5mm} N_\alpha(E_{c.m.})
    \nonumber \\
    &=& f_\beta (E_{c.m.}) \mathop{\sum}_L P_L(E_{c.m.}, a_c)
    \nonumber \\
    & \cdot& \left| \frac{\mathop{\sum}_n \tilde{\gamma}_{n L}B_{n L}/(\tilde{E}_{n L}-E_{c.m.})+{\cal R}_{\beta L}}
    {1-[L_L(k_i a_c)-a_c \tilde{b}_c] R_L(E_{c.m.})} \right|^2\!\!,
    \label{eq:beta}
  \end{eqnarray}
where $B_{n L}$ is the $\beta$-feeding amplitude.
${\cal R}_{\beta L}$ is the energy-independent background.
$f_\beta(E_{c.m.})$ is the integrated Fermi function for the $\beta$-allowed transition, and it is given by
  \begin{eqnarray}
    f_\beta(E_{c.m.}) &=& \int^{Q_\beta}_{m_ec^2} F(Z,E_e) p_e E_e (Q_\beta-E_e)^2 dE_e,
    \nonumber \\
  \end{eqnarray}
where $E_e$ and $p_e$ are the energy and momentum of the emitted electron, respectively.
$F(Z,E_e)$ is the Fermi function;
$Z$ denotes the charge of the daughter nucleus, $Z=8$.
$Q_\beta$ is the Q-value for $\beta$-decay, defined as the mass difference between parent and daughter nuclei, $E_{c.m.}=Q_\beta-E_e$.
$m_ec^2$ is the rest mass of electron.

The value of $B_{1 1}$ for the subthreshold 1$^-_1$ state is given by
  \begin{eqnarray}
    B^2_{1 1} &=& \frac{N_c}{\pi f_\beta(E_{11})}\frac{Y_{1}}{Y_{2}},
    \label{eq:b11}
  \end{eqnarray}
  where $Y_n$ is the $\beta$-decay branching ratio of $^{16}$N to the 1$^-_n$ state: $Y_1=(4.8\pm0.4)\times10^{-2}$, $Y_2=(1.20\pm0.05)\times10^{-5}$ \cite{Til93}.
  $N_c$ is the total number of count.
  Thus, I find $B_{11}/\sqrt{N_c}= (1.30\pm 0.08)$.
  $B_{2 1}$ for 1$^-_2$ and ${\cal R}_{\beta L}$ are the adjustable parameters.
  $B_{3 1}$ and $B_{4 1}$ are set to be zero.

\subsection{Linear approximation of the formal parameters}

The relation between the parameters is obtained from the phase equivalence in the $R$-matrix and Breit-Wigner formula.
From Eqs.~(\ref{eq:rmat}) and (\ref{eq:ps-r}), the phase shift of the single pole is obtained in the $R$-matrix theory as
\begin{eqnarray}
  && \hspace{-5mm}\delta_L^R
  \nonumber \\
  &=& \arctan\left( \frac{\tilde{\Gamma}_{nL}/2}{\tilde{E}_{nL} -E_{c.m.} -\tilde{\gamma}_{nL}^2 [\Delta_L(E_{c.m.}, a_c)-a_c \tilde{b}_c]}\right).
  \nonumber \\
\end{eqnarray}
For the Breit-Wigner formula, the phase shift is given by
\begin{eqnarray}
  \delta_{BW} &=& \arctan\left( \frac{\Gamma_{nL}/2}{E_{nL}-E_{c.m.}}\right).
\end{eqnarray}
So, $\delta_{BW} = \delta_L^R$ is found, if the formal parameters are defined as
\begin{eqnarray}
  \tilde{E}_{n L} &=& E_{n L} + \tilde{\gamma}_{n L}^2 [\Delta_L(E_{n L}, a_c)-a_c \tilde{b}_c ],
  \label{eq:formal-e-app}\\
  \tilde{\gamma}_{n L}^2 &=& \frac{\gamma_{n L}^2}{1-\gamma_{n L}^2 \Delta_L^\prime(E_{n L}, a_c)}.
  \label{eq:formal-g-app}
\end{eqnarray}
To obtain Eqs.~(\ref{eq:formal-e-app}) and (\ref{eq:formal-g-app}), the linear approximation to the shift function is used,
\begin{eqnarray}
  \Delta_L(E_{c.m.}, a_c)
  &\approx& \Delta_L(E_{n L}, a_c)
  \nonumber \\
  &+&(E_{c.m.}-E_{n L})\Delta_L^\prime(E_{n L}, a_c).
  \label{eq:linear}
\end{eqnarray}
In general, Eq.~(\ref{eq:linear}) is assumed to be accurate.
I assess the linear approximation in the present article, including the higher-order terms of expansion.

If the multi-poles should be considered in $R$-matrix, the observed resonance energy is given by $\delta_L^R=\pi/2$, and it satisfies the relation of
\begin{eqnarray}
  [\Delta_L(E_{nL}, a_c)-a_c\tilde{b}_c] R_L(E_{nL}) &=& 1.
\end{eqnarray}

\subsection{Orthogonality of the internal wavefunctions}

The internal wavefunctions $\tilde{\varphi}_{nL}$ are supposed to be orthonormal over the interaction region although they are not numerically obtained.
The $n$th internal wave satisfies 
\begin{eqnarray}
  \Big[ -\frac{\hbar^2}{2\mu} \Big( \frac{d^2}{dr^2}-\frac{L(L+1)}{r^2} \Big) + \hat{V} \,\Big] \,\tilde{\varphi}_{nL}
  &=& \tilde{E}_{nL} \,\tilde{\varphi}_{nL},
  \label{eq:intwv}
  \nonumber \\
\end{eqnarray}
where $\hat{V}$ is interaction.
The $n^\prime$th internal wave satisfies the similar equation.
Subtracting Eq.~(\ref{eq:intwv}) multiplied by $\tilde{\varphi}_{n^\prime L}$ from the exchanged equation, I obtain 
\begin{eqnarray}
  && \hspace{-5mm}
  -\frac{\hbar^2}{2\mu} \Big[ \tilde{\varphi}_{nL}(r)\tilde{\varphi}_{n^\prime L}^{\prime\prime}(r)
    - \tilde{\varphi}_{n^\prime L}(r)\tilde{\varphi}_{nL}^{\prime\prime}(r) \Big]
  \nonumber \\
  &=& (\tilde{E}_{n^\prime L}-\tilde{E}_{nL}) \,\tilde{\varphi}_{nL}(r)\tilde{\varphi}_{n^\prime L}(r).
\end{eqnarray}
If this equation is integrated from $r=0$ to $r=a_c$, I find
\begin{eqnarray}
  \int^{a_c}_0 \tilde{\varphi}_{nL}(r) \tilde{\varphi}_{n^\prime L}(r) dr &=& \delta_{n n^\prime}.
  \label{eq:orth}
\end{eqnarray}
To obtain Eq.~(\ref{eq:orth}), the logarithmic derivative $\tilde{b}_c$ of $n$th state is assumed to be the same as that of $n^\prime$th state,
\begin{eqnarray}
  \tilde{b}_c &\equiv& \frac{\tilde{\varphi}^\prime_{nL}(a_c)}{\tilde{\varphi}_{nL}(a_c)}
  \,\,=\,\, \frac{\tilde{\varphi}^\prime_{n^\prime L}(a_c)}{\tilde{\varphi}_{n^\prime L}(a_c)}.
  \label{eq:bc}
\end{eqnarray}
$\tilde{b}_c$ is an arbitrary constant in the $R$-matrix theory.
Using Eq.~(\ref{eq:formal-g-app}), it is also defined as
\begin{eqnarray}
  \tilde{b}_c
  &=&
    \frac{1}{1-\gamma^2_{nL}\Delta^\prime_L(E_{nL}, a_c)} \bigg[ \,b_{nL}+\frac{\gamma^2_{nL}}{2} \nonumber \\
  &\cdot&
    \left(-\frac{\Delta_L^\prime(E_{nL}, a_c)}{a_c}+\frac{d\Delta_L^\prime(E_{nL}, a_c)}{dr} \right)\bigg],
\end{eqnarray}
where $b_{nL}=\varphi^\prime_{nL}(a_c) /\varphi_{nL}(a_c)$.
$\varphi_{nL}(r)$ is the internal wavefunction for the observed energy $E_{nL}$.
If $\tilde{b}_c=0$ is used, I find
\begin{eqnarray}
  b_{nL} &=&
  \frac{\gamma^2_{nL}}{2}
  \left( \frac{\Delta_L^\prime(E_{nL}, a_c)}{a_c}
  -\frac{d\Delta_L^\prime(E_{nL}, a_c)}{dr} \right).
  \label{eq:bnl}
  \nonumber \\
\end{eqnarray}
$b_{nL}$ is dependent on states and $a_c$.
In addition, $b_{nL}$ depends on energy if the higher-order correction of Eq.~(\ref{eq:q-ecm}) is included.
If $\gamma^2_{nL}\ll 1$, $\tilde{b}_c\approx b_c$ is found, $b_c\equiv b_{nL}\approx b_{n^\prime L}$.

In the present article, I use $\tilde{b}_c=0$.
Therefore, $\tilde{\varphi}_{nL}(r)$ satisfies the orthogonality by the definition.
For $\varphi_{nL}(r)$, $b_c= b_{11}\approx b_{21} \approx 0$ is found in Fig.~\ref{fig:wf} and Eq.~(\ref{eq:bnl}).

  

\end{document}